\documentclass[twocolumn]{aastex62}
\shorttitle{Medium-band Reverberation Mapping}
\shortauthors{Kim et al.}
\begin{document}

\title{Medium-band Photometry Reverberation Mapping of Nearby Active Galactic Nuclei}

\correspondingauthor{Myungshin Im}
\email{mim@astro.snu.ac.kr}

%\author[0000-0002-0786-7307]{Greg J. Schwarz}
\author{Joonho Kim}
\affil{Center for the Exploration of the Origin of the Universe (CEOU),\\
Astronomy Program, Department of Physics and Astronomy,\\
Seoul National University, Seoul 08826, Republic of Korea}

\author{Myungshin Im}
\affil{Center for the Exploration of the Origin of the Universe (CEOU),\\
Astronomy Program, Department of Physics and Astronomy,\\
Seoul National University, Seoul 08826, Republic of Korea}

\author{Changsu Choi}
\affil{Center for the Exploration of the Origin of the Universe (CEOU),\\
Astronomy Program, Department of Physics and Astronomy,\\
Seoul National University, Seoul 08826, Republic of Korea}

\author{Sungyong Hwang}
\affil{Center for the Exploration of the Origin of the Universe (CEOU),\\
Astronomy Program, Department of Physics and Astronomy,\\
Seoul National University, Seoul 08826, Republic of Korea}



\begin{abstract}

Reverberation mapping (RM) is one of the most efficient ways to investigate the broad-line region around the central supermassive black holes of active galactic nuclei (AGNs). A common way of performing the RM is to perform a long term spectroscopic monitoring of AGNs, but the spectroscopic monitoring campaign of a large number of AGNs requires an extensive amount of observing time of medium to large size telescopes. As an alternative way, we present the results of photometric RM with medium-band photometry. As the widths of medium-band filters match well with the widths of AGN broad emission lines, the medium-band observation with small telescopes can be a cost-effective way to perform RM. We monitored five nearby AGNs with available spectroscopic RM results showing days to weeks scale variability. Observations were performed for $\sim$3 months with an average of 3 days cadence using three medium-band filters on a 0.43 m telescope. The time lags between the continuum and the H$\alpha$ emission line light curves are calculated using the JAVELIN software and the discrete correlation function. We find time lags of 1.5-15.9 days for these AGNs, which are consistent with the time lags derived from previous spectroscopic RM measurements. This result demonstrates that even a 0.5 m class telescope can perform RM with medium-bands. Furthermore, we show that RM for tens of thousands AGNs is possible with a dedicated 1 m class telescope.

\end{abstract}



\section{Introduction} \label{sec:intro}

Reverberation mapping (RM) is known to be an important way to investigate the structure and kinematics of the broad-line region (BLR) around the central supermassive black holes of the active galactic nuclei (AGNs; \citealt{1982ApJ...255..149B, 1993PASP..105..247P}). According to the AGN unification model \citep{1993ARA&A..31..473A, 1995PASP..107..803U, 1997iagn.book.....P}, the accretion disk around the central black hole is responsible for the thermal continuum emission in the ultraviolet and optical. On the other hand, the broad emission lines (BEL) arise from the recombination in clouds that are sufficiently far from the central engine (light days to years), and these clouds are called the BLR clouds. When the accretion disk activity varies due to a reason (e.g., increased amount of infalling gas), it takes time for the effect to reach BLR, and hence a time lag in the variation between the continuum and the BEL occurs. Then, the time lag can tell us the distance between the central accretion disk to BLR. Once the BLR size is determined, the masses of the supermassive black holes can be measured from the size and the velocity width of the BEL. RM studies have enabled black hole mass ($M_\mathrm{BH}$) measurements for hundreds of AGNs. From such studies, the correlation between the BLR size and the continuum luminosity has been established, which then resulted in $M_\mathrm{BH}$ estimates for hundreds of thousands of AGNs \citep{2000ApJ...533..631K, 2005ApJ...629...61K, 2007ApJ...659..997K, 2004ApJ...613..682P, 2006ApJ...644..133B, 2009ApJ...705..199B, 2010ApJ...716..993B, 2013ApJ...767..149B, 2010ApJ...721..715D, 2010ApJ...724..386K, 2015ApJS..216...17K, 2012ApJ...755...60G, 2017ApJ...851...21G, 2014ApJ...782...45D, 2015ApJ...806...22D, 2016ApJ...820...27D, 2018ApJ...856....6D, 2015ApJS..217...26B, 2015ApJ...804..138H, 2015ApJ...806..109J, 2016ApJ...818...30S, 2017ApJ...840...97F, 2017ApJ...846...79L, 2018ApJ...865...56L}.

The common way of performing RM is to do a long term spectroscopic monitoring. Through the spectroscopy, one measures the velocity width and the line intensity variation directly from the observed spectra that are taken over the course of months to years. Then, the time lag of the continuum and the line intensity can be directly inferred. However, spectroscopic RM requires an extensive amount of observing time of 1-3 m telescopes \citep{2000ApJ...533..631K, 2009ApJ...705..199B, 2016ApJ...818...30S}, and therefore it is difficult to perform on a large number of AGNs ($>100$).

To overcome the shortcomings of the spectroscopic RM, purely photometric RM has successfully been performed adopting broad-bands only or a combination of broad-, medium-, and narrow-bands \citep{2011A&A...535A..73H, 2012A&A...545A..84P, 2013A&A...552A...1P, 2015A&A...576A..73P, 2012ApJ...747...62C, 2012ApJ...750L..43C, 2012ApJ...756...73E, 2015ApJ...801...45H, 2016ApJ...818..137J, 2016ApJ...819..122Z, 2018ApJ...853..116Z}. Such a technique opens up a possibility of performing the RM of more than thousands of AGNs at once, even using a moderately sized telescopes. In its simplest form, the photometric RM uses two bands or more. One covers the wavelength where a BEL is located, and another covers the wavelength where no strong emission line exists. For example, \citet{2011A&A...535A..73H, 2012A&A...545A..84P, 2013A&A...552A...1P, 2015A&A...576A..73P} observed a few AGNs using sets of a broad-band and a narrow-band. The broad-band is for the continuum flux and the narrow-band traces the BEL. Their time lag measurements agree with results from spectroscopic RM. The use of the narrow-band is advantageous for increasing signal-to-noise ratio (S/N) from an emission line, but a narrow-band can cover a very limited redshift range of the AGN emission lines.

On the other hand, RM with broad-band filters can cover an emission line over a wide range of redshifts. The broad-band based RM has been tried on a few AGNs initially \citep{2012ApJ...747...62C, 2012ApJ...750L..43C, 2012ApJ...756...73E, 2016ApJ...819..122Z}, and more recently to a few hundred quasars in the Sloan Digital Sky Survey (SDSS) Stripe 82 field with $\sim60$ epochs over a $\sim10$ year period \citep{2015ApJ...801...45H}. One problem of the broad-band RM approach is that broad-bands are often too broad to observe only the continuum without contamination by a BEL. Also, the emission lines are often too narrow in comparison to the width of the broad-band filter, and the contrast between the continuum and the line is difficult to measure. \citet{2018ApJ...853..116Z} show that the line flux to continuum flux ratio should be 6 \% or larger for the broad-band RM to be effective. To overcome these problems, \citet{2016ApJ...818..137J} used a combination of broad-band and medium-band with the band widths of 180-264 \AA\ which are several times narrower than typical broad-band filter widths. They observed 13 AGNs at z = 0.2-0.4, and detected time lags in 6 AGNs. The results agree with the BLR size-luminosity relation.

The above examples show the usefulness of a broad-band or medium-band RM approach. However, purely medium-band based RM with multiple medium-bands can be even more powerful. Multiple medium-bands can cover multiple BELs over a wide range of redshifts \citep{2019ApJ...870...86K}. The medium-bands can limit contamination from BELs while sampling a fair amount of continuum flux to trace the continuum variation accurately. The medium-band is matched reasonably well with FWHMs of typical BELs. For example, for an H$\alpha$ with FWHM $= 3000$ km s$^{-1}$ at $z=0.3$, the observed line width corresponds to 85 \AA, comparable to a medium-band width within a factor of a few, which is helpful for increasing the line flux to continuum ratio. The use of a medium-band also opens a possibility of using small telescopes for AGN RM.

To verify the usefulness of the medium-band photometric RM through a comparison with spectroscopic RM results, we used three medium-bands on a 0.43m telescope to monitor five nearby AGNs that have been studied previously with spectroscopic RM. In section 2, we describe the target AGNs, the system used for the observation, and the observation campaign. Section 3 explains the analysis of the data and the light-curve analysis. We estimate the AGN variability based on the light curve in section 4. In Section 5, we present the time lag measurements and how they compare with previous works. Finally, we end the paper with discussion for future prospects and the conclusion of this study.

%%% TABLE %%%%%%%%%%%%%%%%%%%%%%%%%%%%%%%%%%%%%%%%%%%%%%%%%%%%%%%%%%%%%%%%%%%%%
\begin{deluxetable*}{lccccccc}
\tablecaption{List of the monitored AGNs.\label{tab:table}}
\tablehead{
\colhead{Name} & \colhead{R.A.} & \colhead{Decl.} & \colhead{$V$ mag} & \colhead{Redshift} & \colhead{Epochs} & \colhead {Average Time Spacing} & \colhead{Monitoring Start-End} 
}
\startdata
Mrk 1310 & 12:01:14.3 & -03:40:41.1 & 15.89\tablenotemark{a} & 0.019 & 28 & 3.0 days & 2017.05.02 - 07.24\\
NGC 4593 & 12:39:39.5 & -05:20:39.2 & 13.15\tablenotemark{a} & 0.008 & 29 & 3.4 days & 2017.04.30 - 08.07\\
NGC 4748 & 12:52:12.5 & -13:24:52.8 & 14.03\tablenotemark{a} & 0.014 & 33 & 2.8 days & 2017.05.15 - 08.15\\
NGC 6814 & 19:42:40.6 & -10:19:25.5 & 14.21\tablenotemark{a} & 0.005 & 53 & 2.0 days & 2017.05.04 - 09.10\\
NGC 7469 & 23:03:15.7 & +08:52:25.3 & 12.34\tablenotemark{b} & 0.016 & 32 & 3.3 days & 2017.05.11 - 09.02\\
\enddata
\tablenotetext{a}{\citet{veron-cetty2010}}
\tablenotetext{b}{\citet{2007ApJS..173..185G}}
%\tablecomments{Note that {\tt \string \colnumbers} does not work with the vertical line alignment token. If you want vertical lines in the headers you can not use this command at this time.}
\end{deluxetable*}
%%%%%%%%%%%%%%%%%%%%%%%%%%%%%%%%%%%%%%%%%%%%%%%%%%%%%%%%%%%%%%%%%%%%%%%%%%%%%%%

%
\begin{figure*}
\centering
\includegraphics[width=70mm,angle=270]{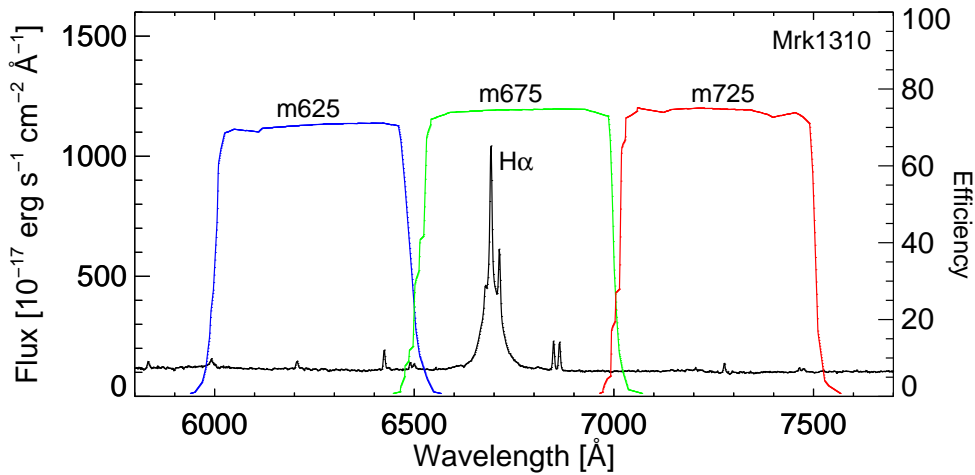}
\caption{SDSS spectrum of Mrk 1310 and transmission curves of the medium-band filters used in this study. As shown here, $m675$ covers the H$\alpha$ emission line and the others cover the continuum.}
\end{figure*}

\section{Sample and Observation} \label{sec:obs}

For the medium-band photometric RM, we used the Lee Sang Gak Telescope (LSGT; \citealt{2015JKAS...48..207I}) and its imaging camera, SNUCAM-II \citep{2017JKAS...50...71C}. LSGT is a 0.43 m diameter telescope located at the Siding Spring Observatory (SSO). The SNUCAM-II camera has a suite of $\sim$50.0 nm width medium-bands covering the wavelength range of 400 to 1050 nm.

For our targets, we selected AGNs that have been studied using the H$\alpha$ emission line for RM at low redshifts ($z < 0.02$), since H$\alpha$ is the strongest emission line and easy to monitor with the small telescope. Additionally, we imposed a requirement that the time lag is less than 2 weeks so that we can obtain results from several months of monitoring campaign. With the above criteria, we select five nearby AGNs, Mrk 1310, NGC 4593, NGC 4748, NGC 6814, and NGC 7469 from \citet{2004ApJ...613..682P, 2014ApJ...795..149P} and \citet{2010ApJ...716..993B}. Table 1 shows the basic properties of our target AGNs, along with the observation summary for each target.

We observed these targets using three medium-bands, $m625$ (596.2-658.7 nm), $m675$ (643.3-712.0 nm), and $m725$ (693.6-754.8 nm). Figure 1 shows the filter transmission curves of these medium-band filters. As shown in the figure, $m675$ covers the H$\alpha$ emission line and the others cover the continuum. We took five frames of 5 minutes exposure time in each band at each epoch. We tried to observe the targets every night. One day cadence observation was possible for patches of extended periods, but because of the weather condition and the Moon distance restriction, the average observational cadence was about 3 days. Typical seeing was $2\farcs0$ to $4\farcs0$. and poor weather condition made it worse to be $\sim5\farcs0$.

The basic data reduction procedure (bias subtraction, dark subtraction, and flat-fielding) was performed with an automatic pipeline as soon as the data were taken. Astrometric solution was found using the Astrometry.net software \citep{2010AJ....139.1782L}, which returned astrometric solutions to a $0\farcs34$ rms accuracy. Then, we used the SWarp software \citep{2002ASPC..281..228B} to combine the five images of 5 minutes exposure to make a single epoch image.

\section{Medium-band Photometry and Light Curve} \label{sec:phot}
When comparing aperture photometry results at different epochs, difference in seeing conditions systematically biases the photometry result. Therefore, we convolved all the epoch images with a Gaussian kernel to make them to have a common seeing FWHM $= 3\farcs8$ to $5\farcs3$ (object-dependent). To exact the flux of the central AGNs, SExtractor \citep{1996A&AS..117..393B} was used with a $12\farcs0$ diameter aperture which corresponds to 2-3 times of the seeing FWHM of the convolved images. Figure 2 shows the images of the five AGNs, and the size of the aperture, demonstrating that the aperture excludes most of the host galaxy light.

\begin{figure*}
\centering
\includegraphics[width=180mm,angle=0]{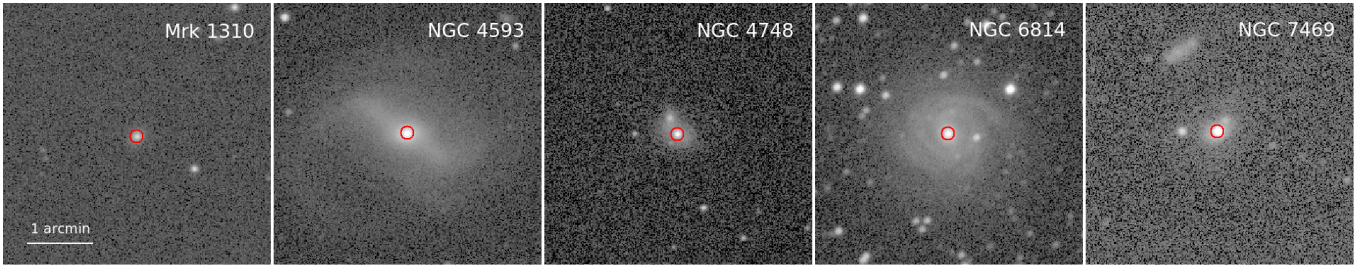}
\caption{The images of the five AGNs in $m675$. The red circle indicates the $12\farcs0$ aperture that we used to measure the AGN flux.}
\end{figure*}
\begin{figure}
\centering
\includegraphics[width=100mm,angle=270]{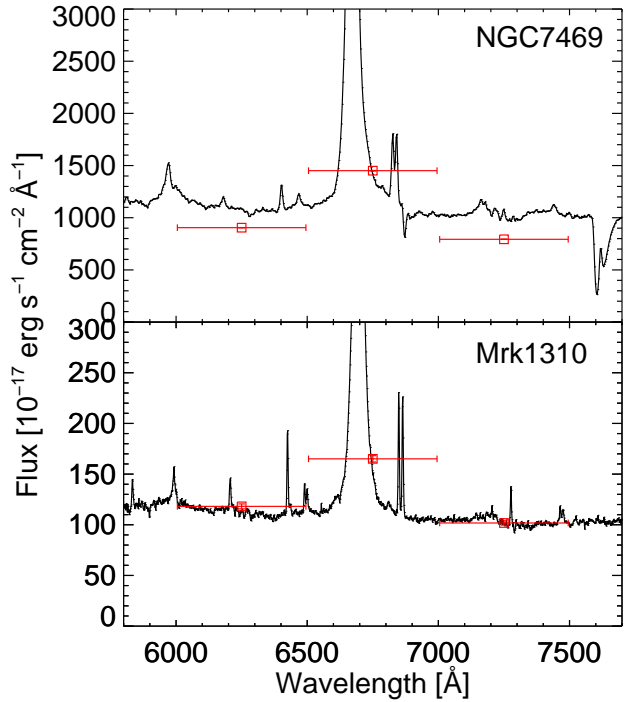}
\caption{Comparison between the spectra and the medium-band photometry (squares with error bars) of NGC 7469 and Mrk 1310. The horizontal line of each point shows the width of the medium-band filter.}
\end{figure}

For the photometric calibration, we used magnitudes of stars in the AAVSO Photometric All-Sky Survey (APASS) catalog \citep{2016yCat.2336....0H}. However, the APASS catalog provides $B,~V,~g,~r,$ and $i$ magnitudes only. To obtain the photometry zero-point of each medium-band image, we first identified photometric reference stars from the APASS catalog as stars that are brighter than 15 mag in each broad-band and within $10\arcmin$ away from the target AGN. Then, we searched for the stellar spectral energy distribution (SED) template among the stellar SED templates of \citet{1983ApJS...52..121G} that best fits the five broad-band magnitudes of each photometry reference star. Only the SEDs that return the reduced $\chi^{2}$ values less than 10 are adopted to derive the medium-band photometry zero-points. Then, applying the transmission curve of each medium-band, we calculated synthetic medium-band magnitudes of the selected stars, and used them for measuring magnitude zero-points (see also \citealt{2017JKAS...50...71C} for more details on this procedure). Typically, 10-70 stars were used to derive the photometric zero-points in each band. The zero-point errors were derived from the rms dispersion of the zero-points from these stars, and a typical zero-point error is found to be 0.01 mag. To verify that the derived medium-band photometry traces the emission line feature in the spectrum well, we compared the photometric result with available spectra of our targets. Figure 3 shows the spectra of NGC 7469 \citep{1995ApJS...98..129K} and Mrk 1310 \citep{2018ApJS..235...42A} along with the medium-band photometry data points. For the medium-band photometry in this case, we used a $3\farcs0$ diameter aperture to be consistent with the aperture size of the SDSS spectrum (for Mrk 1310) or be close to the spatial extraction window of the long-slit spectrum (for NGC 7469). As in the figure, the medium-band photometry traces well the spectral shape around H$\alpha$ for both Mrk 1310 and NGC 7469. There is a small constant offset between the medium-band data and the spectrum for NGC 7469, but this can be attributed to the long term variability of the AGN or a mismatch in the aperture between the medium-band and the long-slit spectrum.
%1.5"x5.8"
%\citet{1983ApJS...52..121G}

%%% TABLE %%%%%%%%%%%%%%%%%%%%%%%%%%%%%%%%%%%%%%%%%%%%%%%%%%%%%%%%%%%%%%%%%%%%%
\begin{deluxetable*}{lccccccc}
\tablecaption{The continuum and the H$\alpha$ emission line light curve of five nearby AGNs.\label{tab:table}}
\tablehead{
\colhead{Name} & \colhead{MJD} & \colhead{$m625$} & \colhead{$m675$} &\colhead{$m725$} & \colhead{Continuum flux} & \colhead{H$\alpha$ flux}\\[-1.5ex]
\colhead{(1)} & \colhead{(2)} & \colhead{(3)} & \colhead{(4)} &\colhead{(5)} & \colhead{(6)} & \colhead{(7)}
}
\startdata
Mrk1310 & 57875.5431 & 15.0886 $\pm$ 0.0079 & 14.7209 $\pm$ 0.0062 & 14.8894 $\pm$ 0.0085 & 0.4269 $\pm$ 0.0053 & 0.1090 $\pm$ 0.0036 \\
Mrk1310 & 57876.3990 & 15.0775 $\pm$ 0.0165 & 14.6971 $\pm$ 0.0116 & 14.9046 $\pm$ 0.0170 & 0.4262 $\pm$ 0.0107 & 0.1234 $\pm$ 0.0069 \\
Mrk1310 & 57884.4938 & 15.1511 $\pm$ 0.0153 & 14.8072 $\pm$ 0.0115 & 14.9636 $\pm$ 0.0138 & 0.4029 $\pm$ 0.0089 & 0.0933 $\pm$ 0.0063 \\
Mrk1310 & 57888.5552 & 15.1283 $\pm$ 0.0110 & 14.7662 $\pm$ 0.0083 & 14.9526 $\pm$ 0.0106 & 0.4106 $\pm$ 0.0068 & 0.1071 $\pm$ 0.0047 \\
Mrk1310 & 57889.5500 & 15.1502 $\pm$ 0.0107 & 14.8038 $\pm$ 0.0156 & 14.9385 $\pm$ 0.0192 & 0.4089 $\pm$ 0.0101 & 0.0880 $\pm$ 0.0086 \\
\enddata
\tablenotetext{}{\textbf{Notes.} (1) AGN name, (2) Modified Julian Date of observation, (3)-(5) AB mag, (6) the 675 nm continuum flux in 10$^{-25}$ erg s$^{-1}$ cm$^{-2}$ Hz$^{-1}$ (observed frame) and its error, (7) H$\alpha$ flux in 10$^{-25}$ erg s$^{-1}$ cm$^{-2}$ Hz$^{-1}$ (observed frame) and its error}
\tablenotetext{}{(This table is published in its entirety in the machine-readable format. A portion is shown here for guidance regarding its form and content.)}
\end{deluxetable*}
%%%%%%%%%%%%%%%%%%%%%%%%%%%%%%%%%%%%%%%%%%%%%%%%%%%%%%%%%%%%%%%%%%%%%%%%%%%%%%%

We employed differential photometry to construct the light curve \citep{2018JKAS...51...89K}. For this, bright stars ($< 15$ mag in all the medium-bands) that are not near the edge of the image are selected as comparison stars. After differential photometry between the comparison stars, those with a relatively large magnitude variation ($>$ 0.01 mag) are removed from the list of the comparison stars. By subtracting the light curves of the AGNs from the light curves of the comparison stars, differential light curves of each AGN are obtained. For the continuum flux at $m675$, we simply take the average of the $m625$ and $m725$ fluxes. Then, the H$\alpha$ flux is obtained by subtracting the continuum flux from the observed $m675$ flux. The continuum and the emission line light curves are shown in Figures 4 and 5 for all the target AGNs and a comparison star. The flux scale of the light curve is set to be the flux at the first epoch of the observation. In Table 2, we provide light curves of five AGNs in machine-readable format.

\section{AGN Variability} \label{sec:var}

We examined the variability of the five AGNs to make sure that they were variable during our observation period. This was done by the $\chi^{2}$ test \citep{2018JKAS...51...89K}. Here, the $\chi^{2}$ value is defined as
\begin{equation}
{\chi}^{2} = \sum_{i} \frac{({m}_{i}-<{m}_{i}>)^{2}}{({\sigma}_{i})^{2}}
\end{equation}
where $i$ denotes each epoch, $m_i$ and $\sigma_{i}$ are the magnitude and its error at the epoch $i$, and $<m_{i}>$ is the magnitude averaged over all the epochs. The results of the $\chi^{2}-$ test, the excess variability (given by $\sqrt{m_{rms}^{2}-<{\sigma}_{i}>^{2}}$, where $m_{rms}$ is the rms of the light-curve magnitudes and $<{\sigma}_{i}>$ is the average magnitude error over all the epochs), and the peak-to-peak (p-to-p) variability in each band are given in Table 3. In the table, the reduced $\chi^{2}$ is indicated, which is $\chi^{2}$ divided by the degree of freedom (the number of data points minus 1). The $\chi^2-$ test result shows that all of the AGNs are variable in all of the three medium-band filters at 99.9\% confidence level.
% {\sigma}_{var} = \sqrt{{{\sigma}_{mag}}^{2}-{{\sigma}_{cal}}^{2}}

%
\begin{figure*}
\centering
\includegraphics[width=230mm,angle=270]{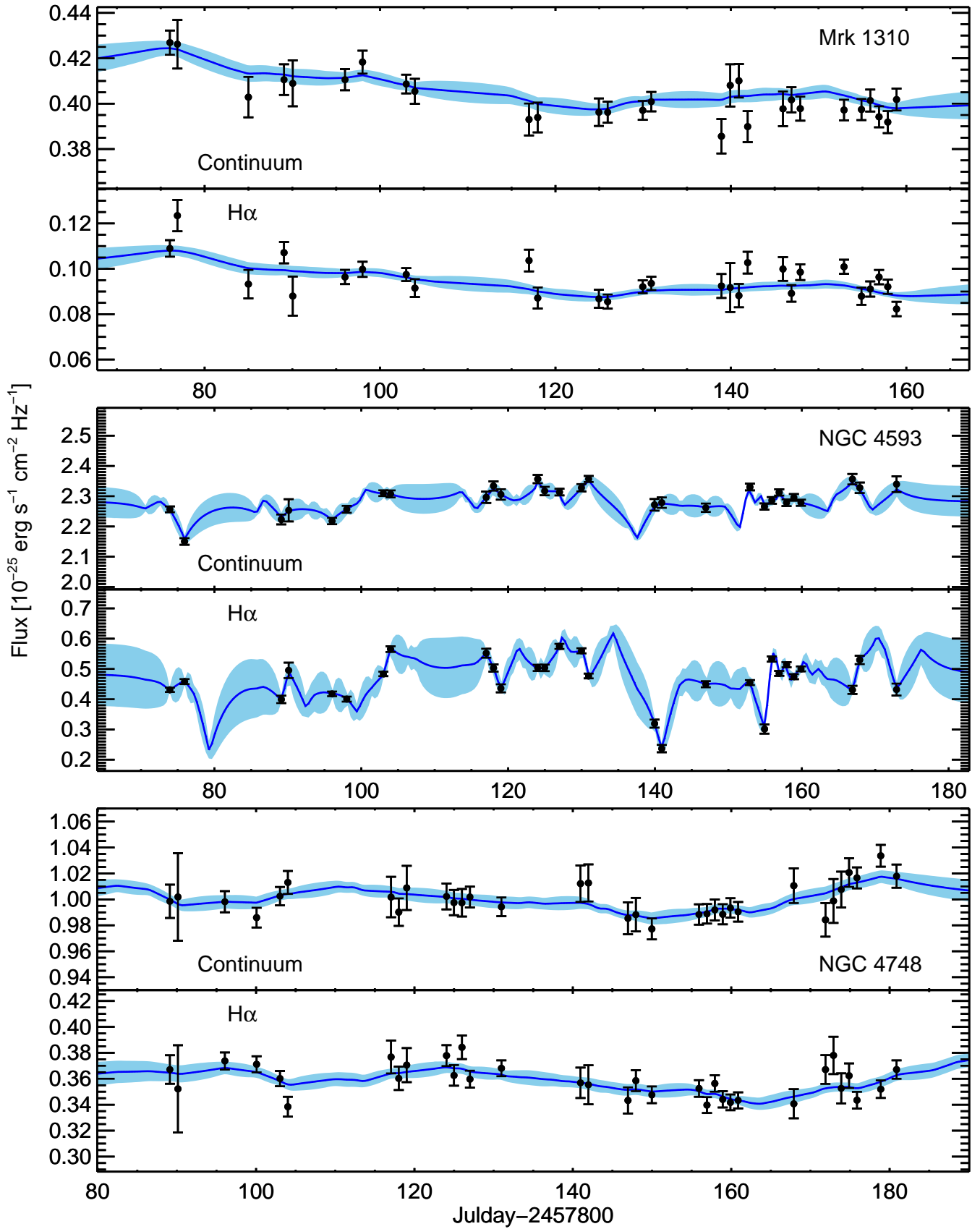}
\caption{The continuum (upper panel) and the H$\alpha$ emission line (lower panel) light curves of five AGNs and of a sample comparison star (the black squares and the error bars). The blue solid lines are the best-fit light curves from JAVELIN, and the filled regions are the uncertainties of the best-fit light curves.}
\end{figure*}
\begin{figure*}
\centering
\includegraphics[width=230mm,angle=270]{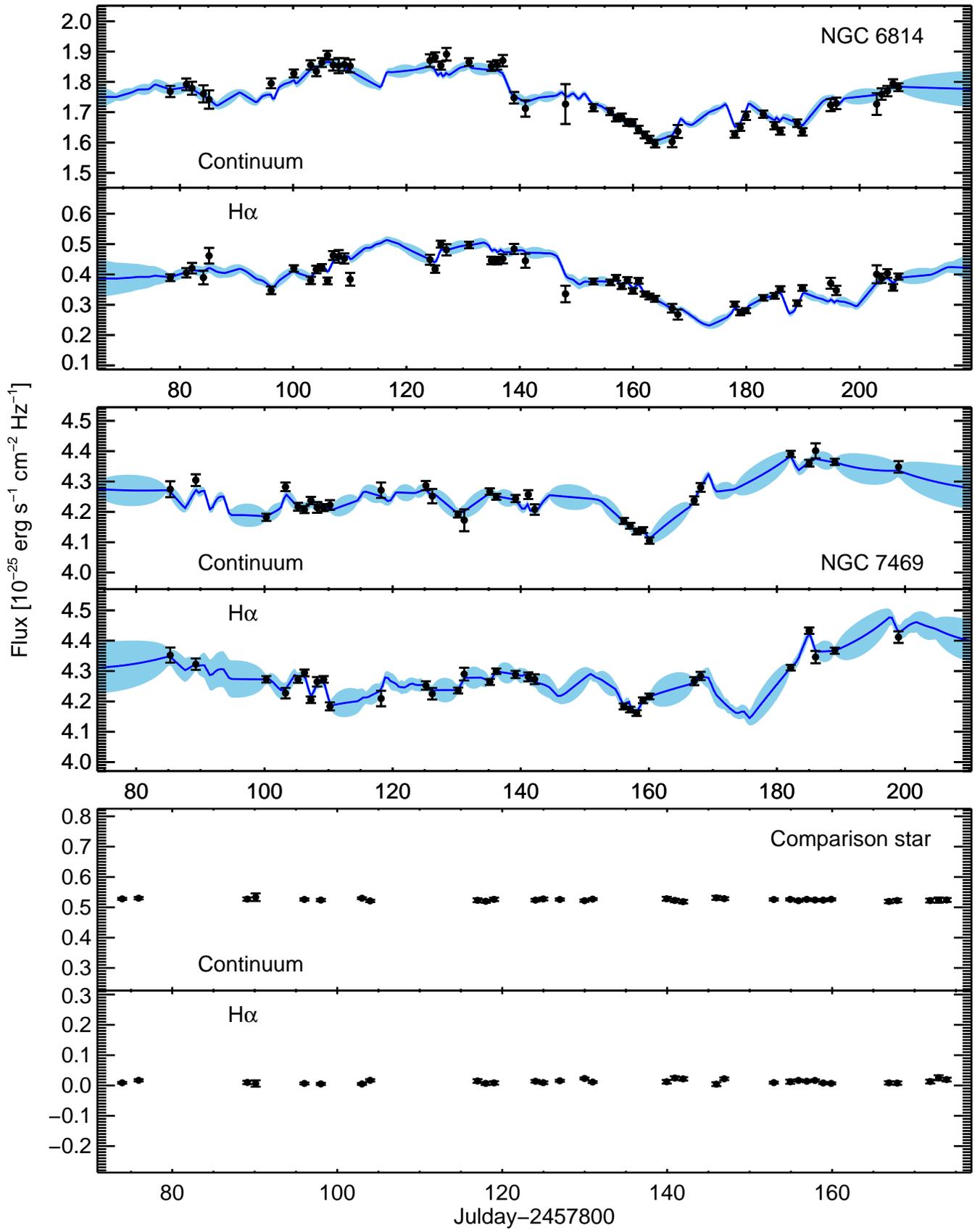}
\caption{Continued.}
\end{figure*}
\begin{figure*}
\centering
\includegraphics[width=230mm,angle=270]{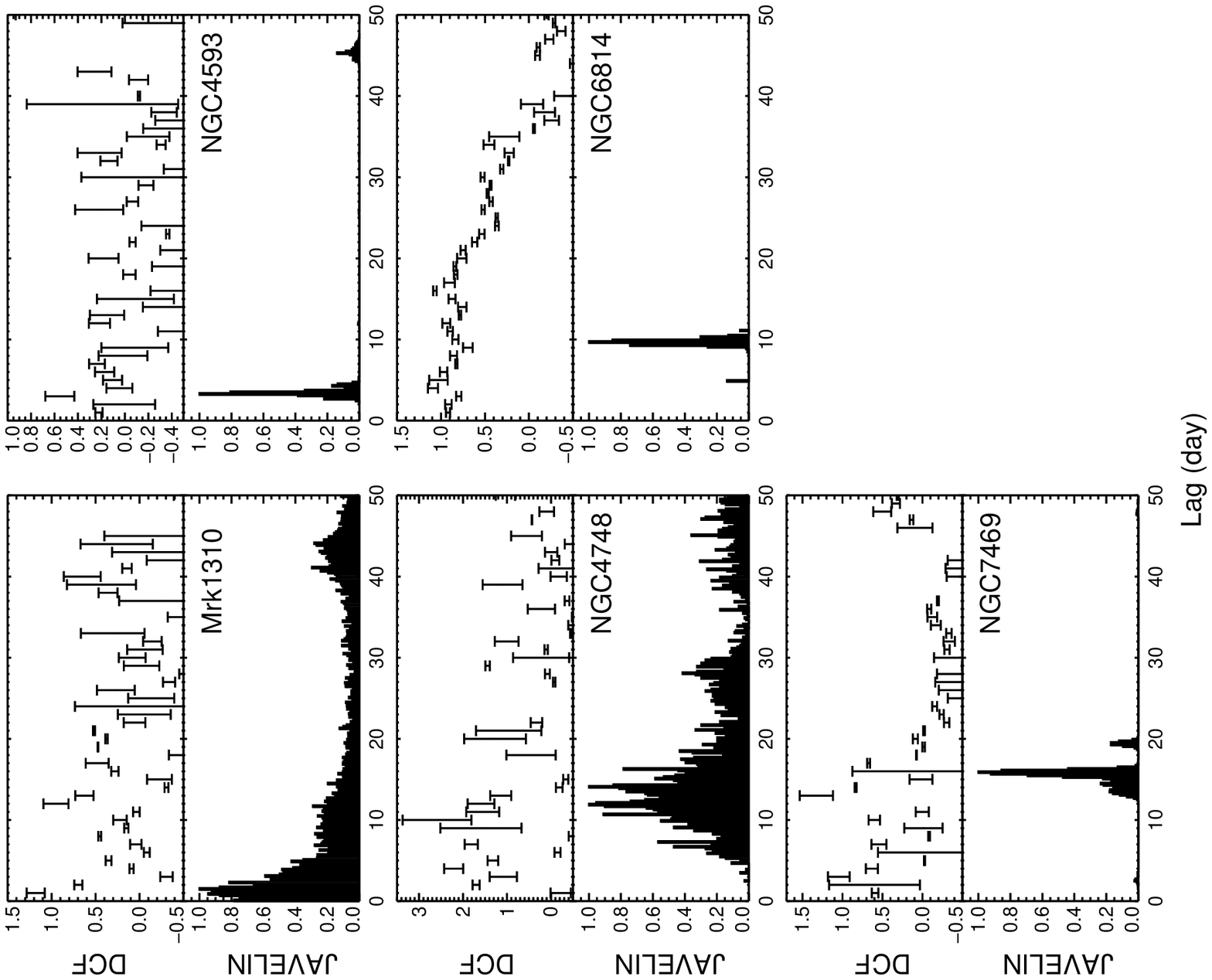}
\caption{Time lag measurement results for the five AGNs in our sample. Upper panels show the DCF values from the DCF analysis, and the lower panels are for the time lag distribution from the JAVELIN analysis.}
\end{figure*}

\section{Result: Medium-band Based Time Lags} \label{sec:result}

We used two different ways to measure the time lag between the continuum and the H$\alpha$ emission line light curves of the AGNs, the JAVELIN software \citep{2011ApJ...735...80Z, 2016ApJ...819..122Z}, and the discrete correlation function (DCF; \citealt{1988ApJ...333..646E}). The JAVELIN software models the continuum light curve using damped random walk \citep{2009ApJ...698..895K} for AGN variability. Then, the continuum light curve is scaled, smoothed and displaced in time and flux to the emission line light curve to find the best-fitting time lag. This procedure is repeated at least 10,000 times by the Markov chain Monte Carlo method to derive many plausible time lags. From the time lag distribution, we identify the time lag that gives the peak value of the distribution as the most probable time lag. The time lag 1$\sigma$ limits are determined to be the upper and lower limits from the most probable time lag value that encompasses 68\% of the time lag distribution. The best-fit light curves for the most probable time lag from JAVELIN are plotted in Figures 4 and 5, along with the observed light curves. Also, the lower panels of Figure 6 show the time lag distribution from JAVELIN for each target with a 0.2 day bin. The peak value of the distribution is normalized to 1.

Another method we tried is DCF, which is developed for analyzing unevenly sampled time-series data. For the DCF analysis, we assumed that the time lag to be 0-50 days, adopting a prior information on the expected time lag. Similarly to the JAVELIN case, many trials are made to search for time lags. The result of the DCF analysis is plotted in the upper panels of Figure 6 for the calculation using a 1 day interval.

Note that we did not detrend the light curve for the above analyses \citep{1999PASP..111.1347W, 2008ApJ...688..837G, 2014ApJ...795..149P}, because the long term variation of our light curves is rather modest.

We find that the measured time lag from two different methods are broadly consistent with each other. However, the JAVELIN results show much clearer peaks than the DCF results. Therefore, we adopt the result from the JAVELIN measurements as the time lag of the H$\alpha$ emission line. In Table 3 and Figure 7, we compare our results with the time lags from previous studies. Also, correlation coefficient between the two time lag measurements is calculated as 0.93, which means our results and the previous results are consistent with each other. Details on the time lag of each target are described below.

\textbf{Mrk 1310}
It has the shortest time lag among five targets. The peak of the DCF is 1 day and the JAVELIN result shows a broad lag distribution. Ignoring the weak second peak and using the lag distribution less than 18 days we find the time lag of $1.5^{+8.2}_{-3.3}$ days. The large error is not surprising, considering our average sampling interval of 3 days. In comparison, \citet{2010ApJ...716..993B} find the time lag between H$\alpha$ and the $V$-band continuum is $4.5^{+0.7}_{-0.6}$ days. Their time lag result with respect to $B$-band continuum is also of order of $4.6^{+0.7}_{-0.6}$ days. Our time lag is consistent with the \citet{2010ApJ...716..993B} result within error.

\textbf{NGC 4593}
For NGC 4593, the DCF shows a peak at 3 days. Similarly, the JAVELIN lag distribution shows a narrow peak at $3.3^{+0.4}_{-0.2}$ days with the very sharp distribution. \citet{2004ApJ...613..682P} found a time lag of $3.2^{+5.6}_{-4.1}$ days, in agreement with our result. Along with NGC 6814, this object shows the strong variability among our sample, and we believe that this is why the JAVELIN distribution has a narrow peak.

\textbf{NGC 4748}
The peak value in the DCF of NGC 4748 is 12 days. From the JAVELIN analysis, we find $11.9^{+5.3}_{-2.6}$ days using the time lag distribution from 0.05 to 22.0 days. It shows the smallest magnitude variations among five AGNs. Contrary to the result of NGC4593, the small variability makes the time lag uncertainty larger (discussed in section 6). However, the result is consistent within uncertainty with the previous measurement of $10.8^{+3.1}_{-3.1}$ days by \citet{2010ApJ...716..993B}.

\textbf{NGC 6814}
Unlike others, the DCF of NGC 6814 shows a clear trend, but a peak seems not to be defined well with a double peak at 4 days and 16 days. On the other hand, JAVELIN lag distribution is sharp with a peak at $9.7^{+0.4}_{-0.4}$ days. It is close to the mean value of double peak in DCF. The time lag of NGC 6814 from \citet{2010ApJ...716..993B} is $9.6^{+2.2}_{-1.7}$ days, in an excellent agreement with our result. Like NGC 4593, the small error in the time lag peak reflects the strongest variation of the light curve (0.06 mag, excess variability).

\textbf{NGC 7469}
The DCF peaks at 13 days and the JAVELIN result shows $15.9^{+0.3}_{-1.7}$ days time lag. Conflicting time lag measurements are presented in previous studies, with $4.9^{+1.7}_{-1.3}$ days time lag of H$\alpha$ emission line \citep{2004ApJ...613..682P}, and $10.9^{+3.5}_{-1.3}$ days using H$\beta$ emission line in a more recent study by \citet{2014ApJ...795..149P}. We note that the H$\alpha$ time lag from our analysis is significantly larger than the values in previous works \citep{2014ApJ...795..149P, 2017MNRAS.466.4759S}. However, considering that the general trend of the H$\alpha$ and H$\beta$ time lag ratio of $\tau$(H$\alpha$):$\tau$(H$\beta$)=1.54:1.00 \citep{2010ApJ...716..993B}, our result is in agreement with the \citet{2014ApJ...795..149P} result. Taking our value and the mass estimators of \citet{2014ApJ...795..149P} and \citet{2015ApJ...806..109J}, the $M_\mathrm{BH}$ estimate from H$\alpha$ is about $4 \times 10^{6}~M_{\odot}$. This value agrees well with the most recent H$\beta$-based $M_\mathrm{BH}$ \citep{2014ApJ...795..149P}. Despite of the small variation of 0.01 mag (excess variability), the time lag measurement has a small error of $\sim$12 \%, comparable to the time lag error of NGC 4593 and NGC 6814, both of which have much larger excess variability of $\sim$0.06 mag. This can be attributed to the fact that NGC 7496 is several times brighter than NGC 4593 and NGC 6814, and thus has the light curve with higher S/N.
%4.2x10^6Msun

%%% TABLE %%%%%%%%%%%%%%%%%%%%%%%%%%%%%%%%%%%%%%%%%%%%%%%%%%%%%%%%%%%%%%%%%%%%%
\begin{deluxetable*}{lcccccccccccc}
\tablecaption{Variability and time lag of monitored AGNs.\label{tab:table}}
\tablehead{
\colhead{Name} & \multicolumn{3}{c}{Excess Variability [mag]} & \multicolumn{3}{c}{p-to-p Variability [mag]} & \multicolumn{3}{c}{Reduced $\chi^{2}$} & \multicolumn{2}{c}{Time Lag [day]} & \colhead{Reference} \\
\ {} & {$m625$} & {$m675$} & {$m725$} & {$m625$} & {$m675$} & {$m725$} & {$m625$} & {$m675$} & {$m725$} & \colhead{This Study} & \colhead{Previous Study} & {}
}
\startdata
Mrk 1310 & ~0.03~ & ~0.04~ & ~0.03~ & ~0.12~ & ~0.16~ & ~0.13~ & 7.4 & 23.1 & 10.6 &$1.5^{+8.2}_{-3.3}$ & $4.5^{+0.7}_{-0.6}$ & 2 \\
NGC 4593 & ~0.04~ & ~0.06~ & ~0.03~ & ~0.17~ & ~0.22~ & ~0.10~ & 141.1 & 257.6 & 57.3 & $3.3^{+0.4}_{-0.2}$ & $3.2^{+5.6}_{-4.1}$ & 1 \\
NGC 4748 & ~0.02~ & ~0.01~ & ~0.01~ & ~0.07~ & ~0.06~ & ~0.06~ & 11.3 & 10.3 & 6.1 & $11.9^{+5.3}_{-2.6}$ & $10.8^{+3.1}_{-3.1}$ & 2 \\
NGC 6814 & ~0.07~ & ~0.08~ & ~0.05~ & ~0.23~ & ~0.28~ & ~0.19~ & 191.8 & 294.3 & 91.6 & $9.7^{+0.4}_{-0.4}$ & $9.6^{+2.2}_{-1.7}$ & 2 \\
NGC 7469 & ~0.03~ & ~0.02~ & ~0.02~ & ~0.11~ & ~0.07~ & ~0.08~ & 289.1 & 232.4 & 143.5 & $15.9^{+0.3}_{-1.7}$ & $4.6^{+1.7}_{-1.3}$ & 1 \\
 & & & & & & & & & & & $10.9^{+3.5}_{-1.3}$ (H$\beta$) & 3 \\
\enddata
\tablenotetext{}{References. (1) \citealt{2004ApJ...613..682P}; (2) \citealt{2010ApJ...716..993B}; (3) \citealt{2014ApJ...795..149P}}
\end{deluxetable*}
%%%%%%%%%%%%%%%%%%%%%%%%%%%%%%%%%%%%%%%%%%%%%%%%%%%%%%%%%%%%%%%%%%%%%%%%%%%%%%%
%Mrk 1310 & ~0.02~ & ~0.03~ & ~0.03~ & ~0.11~ & ~0.15~ & ~0.12~ & 5.5 & 16.4 & 8.6 &$1.5^{+8.2}_{-3.3}$ & $4.5^{+0.7}_{-0.6}$ & 2 \\
%NGC 4593 & ~0.06~ & ~0.05~ & ~0.02~ & ~0.27~ & ~0.26~ & ~0.10~ & 147.5 & 159.1 & 30.8 & $3.3^{+0.4}_{-0.2}$ & $3.2^{+5.6}_{-4.1}$ & 1 \\
%NGC 4748 & ~0.01~ & ~0.01~ & ~0.01~ & ~0.05~ & ~0.06~ & ~0.05~ & 4.0 & 6.0 & 2.1 & $11.9^{+5.3}_{-2.6}$ & $10.8^{+3.1}_{-3.1}$ & 2 \\
%NGC 6814 & ~0.06~ & ~0.06~ & ~0.05~ & ~0.19~ & ~0.24~ & ~0.18~ & 153.0 & 210.5 & 115.0 & $9.7^{+0.4}_{-0.4}$ & $9.6^{+2.2}_{-1.7}$ & 2 \\
%NGC 7469 & ~0.02~ & ~0.01~ & ~0.01~ & ~0.06~ & ~0.04~ & ~0.05~ & 59.0 & 53.8 & 39.6 & $15.9^{+0.3}_{-1.7}$ & $4.6^{+1.7}_{-1.3}$ & 1 \\
%Mrk 1310 & ~0.03~ & ~0.03~ & ~0.03~ & ~0.11~ & ~0.15~ & ~0.12~ & 5.5 & 16.4 & 8.6 &$1.5^{+8.2}_{-3.3}$ & $4.5^{+0.7}_{-0.6}$ & 2 \\
%NGC 4593 & ~0.06~ & ~0.05~ & ~0.02~ & ~0.27~ & ~0.26~ & ~0.10~ & 147.5 & 159.1 & 30.8 & $3.3^{+0.4}_{-0.2}$ & $3.2^{+5.6}_{-4.1}$ & 1 \\
%NGC 4748 & ~0.01~ & ~0.01~ & ~0.01~ & ~0.05~ & ~0.06~ & ~0.05~ & 4.0 & 6.0 & 2.1 & $11.9^{+5.3}_{-2.6}$ & $10.8^{+3.1}_{-3.1}$ & 2 \\
%NGC 6814 & ~0.06~ & ~0.07~ & ~0.06~ & ~0.19~ & ~0.24~ & ~0.18~ & 153.0 & 210.5 & 115.0 & $9.7^{+0.4}_{-0.4}$ & $9.6^{+2.2}_{-1.7}$ & 2 \\
%NGC 7469 & ~0.02~ & ~0.01~ & ~0.01~ & ~0.06~ & ~0.04~ & ~0.05~ & 59.0 & 53.8 & 39.6 & $15.9^{+0.3}_{-1.7}$ & $4.6^{+1.7}_{-1.3}$ & 1 \\

\section{Discussion: Implications for future medium-band RM study} \label{sec:discussion}
The detailed examination of each target in the previous section shows that the accuracy of time lag measurement correlates with the amount of variability as well as S/N, relative cadence, and the number of data points. Obviously, high S/N measurements allow us to examine the time variability better by providing high accuracy photometry. It is easier to catch the variability for highly variable sources, so the time lag accuracy must be improved with the amount of variability. Finally, the number of data points for a given time lag matters, as denser sampling of the light curve can help to understand the overall trend of the variability even if the photometry accuracy is less than the variability. Therefore, we suggest that the product of S/N, the excess variability, and the time lag ($t_{\rm lag}$) divided by the average time spacing of the observation ($\Delta t$) as a proxy of the time lag measurement accuracy. In Figure 8, we show how the time lag uncertainty changes with the product of the excess variability, the average S/N, and $t_{\rm lag}/\Delta t$. Here, the S/N is the average of the measured flux divided by the average of the flux errors.

\begin{figure}
\centering
\includegraphics[width=80mm,angle=270]{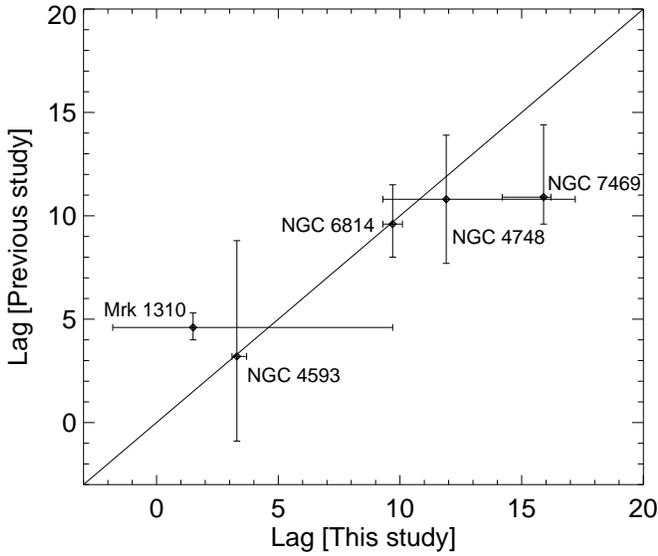}
\caption{Comparison of the time lag results between this study and previous studies. The calculated correlation coefficient of the two time lag measurements is 0.93, meaning that our results and the previous results agree well with each other.}
\end{figure}
\begin{figure}
\centering
\includegraphics[width=80mm,angle=270]{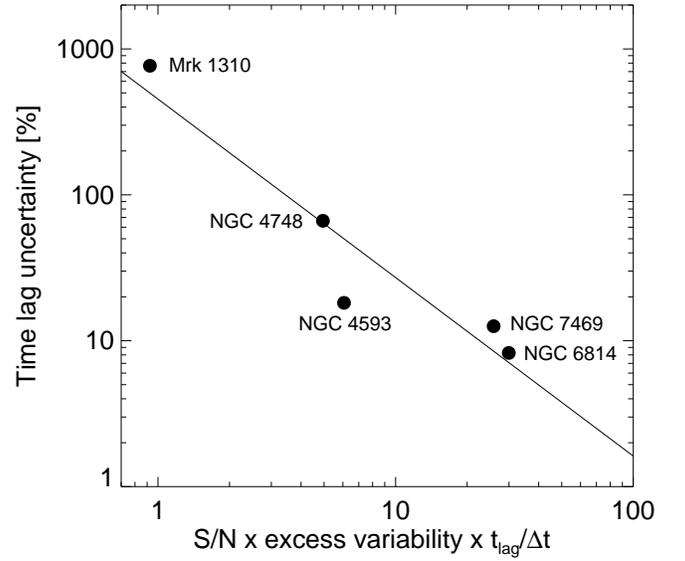}
\caption{A plot showing how the time lag uncertainty decreases with S/N $\times$ excess variability $\times$ $t_{\rm lag}/\Delta t$. The solid line is the fitting result with a slope of -1.22. This shows that the time lag uncertainty is inversely proportional to the S/N, variability, and $t_{\rm lag}/\Delta t$.}
\end{figure}

For a medium-band survey with a photometric accuracy of 10\%, the excess variability of 0.03 mag, and 1 week cadence for an AGN of a 1 year time lag (i.e., $t_{\rm lag}/\Delta t = 52$), Figure 8 shows that the time lag uncertainty is about 10\%. With this result at hand, we can expect how powerful the medium-band RM can be if one uses a dedicated wide-field telescope. For example, a wide-field imaging telescope such as KMTNet \citep{2016JKAS...49...37K} can observe 4 deg$^2$ of the sky at once. If the KMTNet 1.6m telescope is equipped with a suite of medium-band filters similar to those on SNUCAM-II \citep{2017JKAS...50...71C} or SQUEAN \citep{2016PASP..128k5004K}, we expect to be able to reach 22 AB mag (10$\sigma$) at $m675$ with an exposure time of 6 minutes. There are about 300 AGNs at $i < 22$ AB mag \citep{2006AJ....131.2766R} in a single KMTNet field of view. An observation with a set of 8 medium-band filters over 400 - 800 nm (e.g., see \citealt{2017JKAS...50...71C}) can cover BELs of AGNs at various redshifts (e.g., H$\alpha$ at $0 < z < 0.22$, H$\beta$ at $0 < z < 0.65$, MgII at $0.43 < z < 1.86$, and CIV at $1.6 < z < 4.1$), and would take about one hour to complete multi-band imaging of a single field. Therefore, over one night of observation ($\sim 10$ hrs), a wide-field telescope like KMTNet can detect 3000 AGNs in eight medium-bands. If we perform a multi-year, one week cadence RM campaign with such a facility, medium-band RM mapping monitoring observation is possible for $\sim$21,000 AGNs, a substantial increase in the sample size over previous RM studies. We expect to be able to study both low redshift, low luminosity AGNs with short time lags, and high redshift, high luminosity AGNs with long time lags. Since there is a correlation between the AGN luminosity and the time lag, the time lag of 1 week sets the lower limit on the AGN luminosity for the RM to be $L_{bol} = 5 \times 10^{43}$ erg/s. The upper limit on the AGN luminosity will be defined by the length of the monitoring observation as well as the redshift due to time dilation effect. If we adopt a 5 year period monitoring, we can do RM mapping for $L_{bol} = 1.5 \times 10^{48}$ erg/s at $z = 0$ or $L_{bol} = 7 \times 10^{46}$ erg/s at $z = 4$. If we change the monitoring period to 15 years, then the Lbol upper limit can increase. The amount of the AGN variability is anti-correlated with the $L_{bol}$. AGNs with $L_{bol} \sim 10^{47}$ erg/s at $1 < z < 2$ with $\Delta t \sim$ months to years are known to have 0.03 mag variability in $g$-band \citep{2016A&A...585A.129S}, so this roughly sets the upper limit of the $L_{bol}$ of AGNs for this RM study.

\section{Summary} \label{sec:summary}
In this paper, we present the RM results of five AGNs using medium-bands. They are observed over 3-5 months with LSGT, a 0.43m telescope at SSO, and three medium-bands for tracing continuum and H$\alpha$ emission line. All the targets are found to be variable, and the continuum-BEL time lags are calculated. The results show that the time lags from medium-band and spectroscopy agree well within error, and the time lag measurement is possible to 10\% accuracy for objects with 0.03 mag excess variability and S/N=10 for each epoch measurement. Therefore, medium-band photometry is effective to study BLR size. This shows a promise for a dedicated 1 m class, wide-field telescope to perform RM study of tens of thousands AGNs down to $i \sim 22$ mag. Medium-band observations would also open doors to amateur astronomers or institutions with small telescopes performing unique science of the RM study.

\acknowledgments
This work was supported by the National Research Foundation of Korea (NRF) grant, No. 2017R1A3A3001362, funded by
the Korea government (MSIT).

This study used the data taken at the LSGT in Siding Spring Observatory, Australia. We thank Sang Gak Lee for her generous contribution to make the telescope possible. We also thank iTelecope.Net and its staff Brad Moore, Pete Poulos, and Ian Leeder for their management and support for LSGT observation.


\vspace{5mm}
\facilities{LSGT \citep{2015JKAS...48..207I}}


\software{SExtractor \citep{1996A&AS..117..393B}, 
          SWarp \citep{2002ASPC..281..228B}, 
          Astronomy.net \citep{2010AJ....139.1782L}, 
          JAVELIN \citep{2011ApJ...735...80Z, 2016ApJ...819..122Z}
          }

\end{document}